# The Maxwell equations as a Bäcklund transformation

C. J. Papachristou

Department of Physical Sciences, Hellenic Naval Academy, Piraeus, Greece
papachristou@snd.edu.gr

**Abstract**

Bäcklund transformations (BTs) are a useful tool for integrating nonlinear partial differential equations (PDEs). However, the significance of BTs in linear problems should not be ignored. In fact, an important linear system of PDEs in Physics, namely, the Maxwell equations of Electromagnetism, may be viewed as a BT relating the wave equations for the electric and the magnetic field, these equations representing integrability conditions for solution of the Maxwell system. We examine the BT property of this system in detail, both for the vacuum case and for the case of a linear conducting medium.

## 1. Introduction

Bäcklund transformations (BTs) are an effective tool for integrating partial differential equations (PDEs). They are particularly useful for obtaining solutions of nonlinear PDEs, given that these equations are often notoriously hard to solve by direct methods (see [1] and the references therein).

Generally speaking, given two PDEs – say (*a*) and (*b*) – for the unknown functions *u* and *v*, respectively, a BT relating these PDEs is a system of auxiliary PDEs containing both *u* and *v*, such that the consistency (*integrability*) of this system requires that the original PDEs (*a*) and (*b*) be separately satisfied. Then, if a solution of PDE (*a*) is known, a solution of PDE (*b*) is found simply by integrating the BT, without having to integrate the PDE (*b*) directly (which, presumably, is a much harder task).

In addition to being a solution-generating mechanism, BTs may also serve as *recursion operators* for obtaining infinite hierarchies of (generally nonlocal) symmetries and conservation laws of a PDE [1–7]. It is by this method that the full symmetry Lie algebra of the self-dual Yang-Mills equation was found [3,6].

In this article, the nature of which is mostly pedagogical, we adopt a somewhat different (in a sense, inverse) view of a BT, suitable for the treatment of linear problems. Suppose we are given a system of PDEs for the unknown functions *u* and *v*. Suppose, further, that the consistency of this system requires that two PDEs, one for *u* and one for *v*, be separately satisfied (thus, the given system is a BT connecting these PDEs). The PDEs are assumed to possess known solutions for *u* and *v*, each solution depending on a number of parameters. If, by a proper choice of the parameters, these functions are made to satisfy the original differential system, then a solution to this system has been found. In other words, we are seeking solutions of the given system by using known, parameter-dependent solutions of the individual PDEs expressing the integrability conditions of this system. Pairs of functions (*u*,*v*) satisfying the system will be said to represent *BT-conjugate solutions*.

This modified view of the concept of a BT has an important application in Electromagnetism that serves as a paradigm for the significance of BTs in linear problems. As discussed in this paper, the Maxwell equations for a linear medium exactly fit this BT scheme. Indeed, as is well known, the consistency of the Maxwell system requires that the electric and the magnetic field satisfy separate wave equations. These equations have known, parameter-dependent solutions, namely, monochromatic plane waves with arbitrary amplitudes, wave vectors, frequencies, etc. (the "parameters" of the problem). By inserting these solutions into the Maxwell system, one may find the necessary conditions on the parameters in order that the plane waves for the two fields represent BT-conjugate solutions of Maxwell's equations.

The paper is organized as follows:

Section 2 reviews the classical concept of a BT. The solution-generating process by using a BT is demonstrated in a number of examples.

In Sec. 3 the concept of parametric, BT-conjugate solutions is introduced. A simple example illustrates the idea.

In Sec. 4 the Maxwell equations in empty space are shown to constitute a BT in the sense described in Sec. 3. For completeness of presentation (and for the benefit of the student) the process of constructing BT-conjugate plane-wave solutions is presented in detail. This process is, of course, a familiar problem of Electrodynamics but is seen here under a new perspective by using the concept of a BT.

Finally, in Sec. 5 the Maxwell system for a linear conducting medium is similarly examined.

## 2. Bäcklund transformations: definition and examples

The general idea of a Bäcklund transformation (BT) was explained in [1] (see also the references therein). Let us review the main points:

We consider two PDEs $P[u]=0$ and $Q[v]=0$, where the expressions $P[u]$ and $Q[v]$ may contain the unknown functions *u* and *v*, respectively, as well as some of their partial derivatives with respect to the independent variables. For simplicity, we assume that *u* and *v* are functions of only two variables $x$, $t$. Partial derivatives with respect to these variables will be denoted by using subscripts, e.g., $u_x$, $u_t$, $u_{xx}$, $u_{tt}$, $u_{xt}$, etc.

We also consider a system of coupled PDEs for $u$ and $v$,

$$B_i[u,v] = 0, \quad i = 1, 2 \qquad (1)$$

where the expressions $B_i[u,v]$ may contain $u$, $v$ and certain of their partial derivatives with respect to $x$ and $t$. The system (1) is assumed to be integrable for $v$ (the two equations are compatible with each other for solution for $v$) when $u$ satisfies the PDE $P[u]=0$. The solution $v$, then, satisfies the PDE $Q[v]=0$. Conversely, the system (1) is integrable for $u$ if $v$ satisfies the PDE $Q[v]=0$, the solution $u$ then satisfying $P[u]=0$.

If the above assumptions are valid, we say that the system (1) constitutes a BT connecting solutions of $P[u]=0$ with solutions of $Q[v]=0$. In the special case where $P \equiv Q$, i.e., when $u$ and $v$ satisfy the same PDE, the system (1) is called an *auto-Bäcklund* transformation (auto-BT).

Suppose now that we seek solutions of the PDE $P[u]=0$. Also, assume that we possess a BT connecting solutions $u$ of this equation with solutions $v$ of the PDE $Q[v]=0$ (if $P \equiv Q$ the auto-BT connects solutions $u$ and $v$ of the same PDE). Let $v=v_0(x,t)$ be a known solution of $Q[v]=0$. The BT is then a system of equations for the unknown $u$:

$$B_i[u,v_0] = 0, \quad i = 1, 2 \qquad (2)$$

Given that $Q[v_0]=0$, the system (2) is integrable for $u$ and its solution satisfies the PDE $P[u]=0$. We may thus find a solution $u(x,t)$ of $P[u]=0$ without solving the equation itself, simply by integrating the BT (2) with respect to $u$. Of course, the use of this method is meaningful provided that we know a solution $v_0(x,t)$ of $Q[v]=0$ beforehand, as well as that integrating the system (2) for $u$ is simpler than integrating the PDE $P[u]=0$ directly. If the transformation (2) is an auto-BT, then, starting with a known solution $v_0(x,t)$ of $P[u]=0$ and integrating the system (2), we find another solution $u(x,t)$ of the same equation.

Let us see some examples of using a BT to generate solutions of a PDE:

1. The *Cauchy-Riemann relations* of Complex Analysis,

$$u_x = v_y \quad (a) \qquad u_y = -v_x \quad (b) \qquad (3)$$

(here, the variable $t$ has been renamed $y$) constitute an auto-BT for the (linear) *Laplace equation*,

$$P[w] \equiv w_{xx} + w_{yy} = 0 \qquad (4)$$

Indeed, differentiating (3a) with respect to $y$ and (3b) with respect to $x$, and demanding that the *integrability condition* $(u_x)_y=(u_y)_x$ be satisfied, we eliminate the variable $u$ to find the consistency condition that must be obeyed by $v(x,y)$ in order that the system (3) be integrable for $u$:

$$P[v] \equiv v_{xx} + v_{yy} = 0 .$$

Conversely, eliminating $v$ from the system (3) by using the integrability condition $(v_x)_y=(v_y)_x$, we find the necessary condition for $u$ in order for the system to be integrable for $v$:

$$P[u] \equiv u_{xx} + u_{yy} = 0 .$$

Now, let $v_0(x,y)$ be a known solution of the Laplace equation (4). Substituting $v=v_0$ in the system (3), we can integrate the latter with respect to $u$ to find another solution of the Laplace equation. For example, by choosing $v_0(x,y)=xy$ we find the solution $u(x,y)= (x^2 -y^2)/2 +C$.

2. The *Liouville equation* is written

$$P[u] \equiv u_{xt} - e^u = 0 \quad \Leftrightarrow \quad u_{xt} = e^u \qquad (5)$$

Solving the PDE (5) directly is a difficult task in view of this equation's nonlinearity. A solution can be found, however, by using a BT. We thus consider an auxiliary function $v(x,t)$ and an associated linear PDE,

$$Q[v] \equiv v_{xt} = 0 \qquad (6)$$

We also consider the system of first-order PDEs,

$$\begin{aligned} u_x + v_x &= \sqrt{2}\ e^{(u-v)/2} \\ u_t - v_t &= \sqrt{2}\ e^{(u+v)/2} \end{aligned} \qquad (7)$$

It can be shown that the self-consistency of the system (7) requires that $u$ and $v$ independently satisfy the PDEs (5) and (6), respectively. Thus, this system constitutes a BT connecting solutions of (5) and (6). Starting with the trivial solution $v=0$ of (6) and integrating the system

$$u_x = \sqrt{2}\ e^{u/2}, \qquad u_t = \sqrt{2}\ e^{u/2},$$

we find a solution of (5):

$$u(x,t) = -2\ln\left(C - \frac{x+t}{\sqrt{2}}\right).$$

3. The "*sine-Gordon*" *equation* has applications in various areas of Physics, such as in the study of crystalline solids, in the transmission of elastic waves, in Magnetism, in elementary-particle models, etc. The equation (whose name is a pun on the related linear Klein-Gordon equation) is written

$$u_{xt} = \sin u \qquad (8)$$

As can be proven, the differential system

$$\frac{1}{2}(u+v)_x = a\,\sin\left(\frac{u-v}{2}\right)$$
$$\frac{1}{2}(u-v)_t = \frac{1}{a}\,\sin\left(\frac{u+v}{2}\right) \tag{9}$$

[where $a$ ($\neq 0$) is an arbitrary real constant] is a parametric auto-BT for the PDE (8). Starting with the trivial solution $v=0$ of $v_{xt}=\sin v$, and integrating the system

$$u_x = 2a\,\sin\frac{u}{2}\,, \qquad u_t = \frac{2}{a}\,\sin\frac{u}{2}\,,$$

we obtain a new solution of (8):

$$u(x,t) = 4\arctan\left\{C\exp\left(ax+\frac{t}{a}\right)\right\}\,.$$

### 3. BT-conjugate solutions

Consider a system of coupled PDEs for the functions $u$ and $v$ of two independent variables $x, y$:

$$B_i[u,v] = 0\,, \quad i=1,2 \tag{10}$$

Assume that the integrability of this system for both $u$ and $v$ requires that the following PDEs be independently satisfied:

$$P[u]=0 \quad (a) \qquad Q[v]=0 \quad (b) \tag{11}$$

That is, the system (10) represents a BT connecting the PDEs (11). Assume, further, that the PDEs (11) possess parameter-dependent solutions of the form

$$u = f(x,y;\alpha,\beta,\gamma,\ldots)\,,$$
$$v = g(x,y;\kappa,\lambda,\mu,\ldots) \tag{12}$$

where $\alpha$, $\beta$, $\kappa$, $\lambda$, etc., are (real or complex) parameters. If values of these parameters can be determined for which $u$ and $v$ satisfy the system (10), we say that the solutions $u$ and $v$ of the PDEs (11a) and (11b), respectively, are *conjugate through the BT* (10) (or *BT-conjugate*, for short).

Let us see an example: Going back to the Cauchy-Riemann relations (3), we try the following parametric solutions of the Laplace equation (4):

$$u(x,y) = \alpha(x^2-y^2) + \beta x + \gamma y\,,$$
$$v(x,y) = \kappa xy + \lambda x + \mu y\,.$$

Substituting these into the BT (3), we find that $\kappa=2\alpha$, $\mu=\beta$ and $\lambda=-\gamma$. Therefore, the solutions

$$u(x,y) = \alpha(x^2-y^2) + \beta x + \gamma y\,,$$
$$v(x,y) = 2\alpha xy - \gamma x + \beta y$$

of the Laplace equation are BT-conjugate through the Cauchy-Riemann relations.

As a counter-example, let us try a different combination:

$$u(x,y) = \alpha xy\,, \quad v(x,y) = \beta xy\,.$$

Inserting these into the system (3) and taking into account the independence of $x$ and $y$, we find that the only possible values of the parameters $\alpha$ and $\beta$ are $\alpha=\beta=0$, so that $u(x,y)=v(x,y)=0$. Thus, no non-trivial BT-conjugate solutions exist in this case.

### 4. Application to the Maxwell equations in empty space

Maxwell's discovery that all electromagnetic (e/m) disturbances propagate in space as waves running at the speed of light constitutes one of the greatest triumphs of Theoretical Physics of all time. It could justly be said that this discovery is no less important than the Theory of Relativity! It is interesting, from the mathematical point of view, that the wave equations for the electric and the magnetic field are connected to each other through the Maxwell equations in much the same way two PDEs are connected via a Bäcklund transformation. In fact, certain parameter-dependent solutions of the two wave equations are BT-conjugate through the Maxwell system.

In empty space, where no charges or currents (whether free or bound) exist, the Maxwell equations are written in S.I. units [8]:

$$(a)\ \vec{\nabla}\cdot\vec{E} = 0 \qquad (c)\ \vec{\nabla}\times\vec{E} = -\frac{\partial\vec{B}}{\partial t}$$
$$(b)\ \vec{\nabla}\cdot\vec{B} = 0 \qquad (d)\ \vec{\nabla}\times\vec{B} = \varepsilon_0\mu_0\frac{\partial\vec{E}}{\partial t} \tag{13}$$

where $\vec{E}$ and $\vec{B}$ are the electric and the magnetic field, respectively. In order that this system of PDEs be self-consistent (thus integrable for the two fields), certain consistency conditions (or *integrability conditions*) must be satisfied. Four are satisfied automatically:

$$\vec{\nabla}\cdot(\vec{\nabla}\times\vec{E}) = 0\,, \quad \vec{\nabla}\cdot(\vec{\nabla}\times\vec{B}) = 0\,,$$
$$(\vec{\nabla}\cdot\vec{E})_t = \vec{\nabla}\cdot\vec{E}_t\,, \quad (\vec{\nabla}\cdot\vec{B})_t = \vec{\nabla}\cdot\vec{B}_t\,.$$

Two others read:

$$\vec{\nabla}\times(\vec{\nabla}\times\vec{E}) = \vec{\nabla}(\vec{\nabla}\cdot\vec{E}) - \nabla^2\vec{E} \tag{14}$$

$$\vec{\nabla} \times (\vec{\nabla} \times \vec{B}) = \vec{\nabla}(\vec{\nabla} \cdot \vec{B}) - \nabla^2 \vec{B} \qquad (15)$$

Taking the *rot* of (13*c*) and using (14), (13*a*) and (13*d*), we find:

$$\nabla^2 \vec{E} - \varepsilon_0 \mu_0 \frac{\partial^2 \vec{E}}{\partial t^2} = 0 \qquad (16)$$

Similarly, taking the *rot* of (13*d*) and using (15), (13*b*) and (13*c*), we get:

$$\nabla^2 \vec{B} - \varepsilon_0 \mu_0 \frac{\partial^2 \vec{B}}{\partial t^2} = 0 \qquad (17)$$

No new information is furnished by the remaining two integrability conditions,

$$(\vec{\nabla} \times \vec{E})_t = \vec{\nabla} \times \vec{E}_t \;, \quad (\vec{\nabla} \times \vec{B})_t = \vec{\nabla} \times \vec{B}_t \;.$$

Putting

$$\varepsilon_0 \mu_0 \equiv \frac{1}{c^2} \;\Leftrightarrow\; c = \frac{1}{\sqrt{\varepsilon_0 \mu_0}} \qquad (18)$$

we rewrite Eqs. (16) and (17) in wave-equation form:

$$\nabla^2 \vec{E} - \frac{1}{c^2} \frac{\partial^2 \vec{E}}{\partial t^2} = 0 \qquad (19)$$

$$\nabla^2 \vec{B} - \frac{1}{c^2} \frac{\partial^2 \vec{B}}{\partial t^2} = 0 \qquad (20)$$

The PDEs (19) and (20) are consistency conditions that must be separately satisfied by $\vec{E}$ and $\vec{B}$ in order that the differential system (13) be integrable for either field, given the value of the other field. In other words, the system (13) is a BT relating solutions of the wave equations (19) and (20).

It should be noted carefully that the BT (13) is *not* an *auto*-BT! Indeed, although the PDEs (19) and (20) look similar, they concern *different* fields with different physical dimensions and physical properties. A true auto-BT should connect similar objects (such as, e.g., different mathematical expressions for the electric field).

The above wave equations admit plane-wave solutions of the form $\vec{F}(\vec{k} \cdot \vec{r} - \omega t)$, with

$$\frac{\omega}{k} = c \quad \text{where} \quad k = |\vec{k}| \qquad (21)$$

The simplest such solutions are *monochromatic plane waves* of angular frequency $\omega$, propagating in the direction of the wave vector $\vec{k}$:

$$\begin{aligned}\vec{E}(\vec{r},t) &= \vec{E}_0 \exp\{i(\vec{k} \cdot \vec{r} - \omega t)\} \quad (a) \\ \vec{B}(\vec{r},t) &= \vec{B}_0 \exp\{i(\vec{k} \cdot \vec{r} - \omega t)\} \quad (b)\end{aligned} \qquad (22)$$

where the $\vec{E}_0$ and $\vec{B}_0$ represent constant complex amplitudes. Since all constants appearing in equations (22) (that is, amplitudes, frequency and wave vector) can be arbitrarily chosen, they can be regarded as *parameters* on which the solutions (22) of the wave equations depend.

Clearly, although every pair of fields $(\vec{E}, \vec{B})$ that satisfies the Maxwell equations (13) also satisfies the respective wave equations (19) and (20), the converse is not true. This means that the solutions (22) of the wave equation are not *a priori* solutions of the Maxwell system of equations (i.e., do not represent e/m fields). This problem can be remedied, however, by appropriate choice of the parameters. To this end, we substitute the general solutions (22) into the system (13) in order to find the extra conditions this system requires; that is, in order to make the two functions in (22) BT-conjugate solutions of the respective wave equations (19) and (20).

Substituting (22*a*) and (22*b*) into (13*a*) and (13*b*), respectively, and taking into account that $\vec{\nabla} e^{i\vec{k}\cdot\vec{r}} = i\vec{k}\, e^{i\vec{k}\cdot\vec{r}}$, we have:

$$(\vec{E}_0 e^{-i\omega t}) \cdot \vec{\nabla} e^{i\vec{k}\cdot\vec{r}} = 0 \;\Rightarrow\; (\vec{k} \cdot \vec{E}_0)\, e^{i(\vec{k}\cdot\vec{r}-\omega t)} = 0,$$

$$(\vec{B}_0 e^{-i\omega t}) \cdot \vec{\nabla} e^{i\vec{k}\cdot\vec{r}} = 0 \;\Rightarrow\; (\vec{k} \cdot \vec{B}_0)\, e^{i(\vec{k}\cdot\vec{r}-\omega t)} = 0,$$

so that

$$\vec{k} \cdot \vec{E}_0 = 0 \;, \quad \vec{k} \cdot \vec{B}_0 = 0 \,. \qquad (23)$$

Physically, this means that the monochromatic plane e/m wave is a *transverse* wave.

Next, substituting (22*a*) and (22*b*) into (13*c*) and (13*d*), we find:

$$e^{-i\omega t} (\vec{\nabla} e^{i\vec{k}\cdot\vec{r}}) \times \vec{E}_0 = i\omega \vec{B}_0\, e^{i(\vec{k}\cdot\vec{r}-\omega t)} \;\Rightarrow$$

$$(\vec{k} \times \vec{E}_0)\, e^{i(\vec{k}\cdot\vec{r}-\omega t)} = \omega \vec{B}_0\, e^{i(\vec{k}\cdot\vec{r}-\omega t)},$$

$$e^{-i\omega t} (\vec{\nabla} e^{i\vec{k}\cdot\vec{r}}) \times \vec{B}_0 = -i\omega \varepsilon_0 \mu_0 \vec{E}_0\, e^{i(\vec{k}\cdot\vec{r}-\omega t)} \;\Rightarrow$$

$$(\vec{k} \times \vec{B}_0)\, e^{i(\vec{k}\cdot\vec{r}-\omega t)} = -\frac{\omega}{c^2} \vec{E}_0\, e^{i(\vec{k}\cdot\vec{r}-\omega t)},$$

so that

$$\vec{k} \times \vec{E}_0 = \omega \vec{B}_0 \;, \quad \vec{k} \times \vec{B}_0 = -\frac{\omega}{c^2} \vec{E}_0 \qquad (24)$$

This means that the fields $\vec{E}$ and $\vec{B}$ are normal to each other as well as being normal to the direction of propagation. It can be seen that the two vector equations in (24) are not independent of each other; indeed, cross-multiplying the first relation by $\vec{k}$ we get the second one.

Introducing a unit vector $\hat{\tau}$ in the direction of the wave vector $\vec{k}$,

$$\hat{\tau} = \vec{k}/k \quad (k = |\vec{k}| = \omega/c) ,$$

we rewrite the first of Eqs. (24) as

$$\vec{B}_0 = \frac{k}{\omega}(\hat{\tau} \times \vec{E}_0) = \frac{1}{c}(\hat{\tau} \times \vec{E}_0) .$$

The BT-conjugate solutions in (22) are now written:

$$\vec{E}(\vec{r},t) = \vec{E}_0 \exp\{i(\vec{k}\cdot\vec{r} - \omega t)\} ,$$
$$\vec{B}(\vec{r},t) = \frac{1}{c}(\hat{\tau} \times \vec{E}_0)\exp\{i(\vec{k}\cdot\vec{r} - \omega t)\} \quad (25)$$
$$= \frac{1}{c}\hat{\tau} \times \vec{E}$$

As constructed, the complex vector fields in (25) satisfy the Maxwell system (13), which is a homogeneous linear system with real coefficients. Evidently, the real parts of these fields also satisfy this system. To find the expressions for the real solutions (which, after all, carry the physics of the situation) we take the simplest case of a linearly polarized e/m wave and write:

$$\vec{E}_0 = \vec{E}_{0,R}\, e^{i\alpha} \quad (26)$$

where the vector $\vec{E}_{0,R}$ and the number $\alpha$ are real. The *real* versions of the fields (25), then, read:

$$\vec{E} = \vec{E}_{0,R}\cos(\vec{k}\cdot\vec{r} - \omega t + \alpha) ,$$
$$\vec{B} = \frac{1}{c}(\hat{\tau} \times \vec{E}_{0,R})\cos(\vec{k}\cdot\vec{r} - \omega t + \alpha) \quad (27)$$
$$= \frac{1}{c}\hat{\tau} \times \vec{E}$$

We note, in particular, that the fields $\vec{E}$ and $\vec{B}$ "oscillate" in phase.

Our results for the Maxwell equations in vacuum can be extended to the case of a *linear non-conducting medium* upon replacement of $\varepsilon_0$ and $\mu_0$ with $\varepsilon$ and $\mu$, respectively. The speed of propagation of the e/m wave is, in this case,

$$\upsilon = \omega/k = 1/\sqrt{\varepsilon\mu} .$$

## 5. The Maxwell system for a linear conducting medium

In a linear conducting medium of conductivity $\sigma$, in which Ohm's law is satisfied, $\vec{J}_f = \sigma\vec{E}$ (where $\vec{J}_f$ is the free current density), the Maxwell equations read [8]:

$$(a)\ \vec{\nabla}\cdot\vec{E} = 0 \quad (c)\ \vec{\nabla}\times\vec{E} = -\frac{\partial\vec{B}}{\partial t}$$
$$(b)\ \vec{\nabla}\cdot\vec{B} = 0 \quad (d)\ \vec{\nabla}\times\vec{B} = \mu\sigma\vec{E} + \varepsilon\mu\frac{\partial\vec{E}}{\partial t} \quad (28)$$

By the integrability conditions

$$\vec{\nabla}\times(\vec{\nabla}\times\vec{E}) = \vec{\nabla}(\vec{\nabla}\cdot\vec{E}) - \nabla^2\vec{E} ,$$
$$\vec{\nabla}\times(\vec{\nabla}\times\vec{B}) = \vec{\nabla}(\vec{\nabla}\cdot\vec{B}) - \nabla^2\vec{B} ,$$

we get the *modified wave equations*

$$\nabla^2\vec{E} - \varepsilon\mu\frac{\partial^2\vec{E}}{\partial t^2} - \mu\sigma\frac{\partial\vec{E}}{\partial t} = 0$$
$$\nabla^2\vec{B} - \varepsilon\mu\frac{\partial^2\vec{B}}{\partial t^2} - \mu\sigma\frac{\partial\vec{B}}{\partial t} = 0 \quad (29)$$

No new information is furnished by the remaining integrability conditions (cf. Sec. 4).

We observe that the linear differential system (28) is a BT relating solutions of the wave equations (29) (as explained in the previous section, this BT is *not* an auto-BT). As in the vacuum case, we seek BT-conjugate such solutions. As can be verified by direct substitution into Eqs. (29), these PDEs admit parametric plane-wave solutions of the form

$$\vec{E}(\vec{r},t) = \vec{E}_0\exp\{-s\hat{\tau}\cdot\vec{r} + i(\vec{k}\cdot\vec{r} - \omega t)\}$$
$$= \vec{E}_0\exp\left\{\left(i - \frac{s}{k}\right)\vec{k}\cdot\vec{r}\right\}\exp(-i\omega t) ,$$
$$\vec{B}(\vec{r},t) = \vec{B}_0\exp\{-s\hat{\tau}\cdot\vec{r} + i(\vec{k}\cdot\vec{r} - \omega t)\} \quad (30)$$
$$= \vec{B}_0\exp\left\{\left(i - \frac{s}{k}\right)\vec{k}\cdot\vec{r}\right\}\exp(-i\omega t)$$

where $\hat{\tau}$ is the unit vector in the direction of the wave vector $\vec{k}$,

$$\hat{\tau} = \vec{k}/k \quad (k = |\vec{k}| = \omega/\upsilon)$$

($\upsilon$ is the speed of propagation of the wave inside the conducting medium) and where, for given physical characteris-tics $\varepsilon$, $\mu$, $\sigma$ of the medium, the parameters $s$, $k$ and $\omega$ satisfy the algebraic system

$$s^2 - k^2 + \varepsilon\mu\omega^2 = 0 ,$$
$$\mu\sigma\omega - 2sk = 0 \qquad (31)$$

Up to this point the complex amplitudes $\vec{E}_0$ and $\vec{B}_0$ in relations (30) are arbitrary and the vector fields (30) are not *a priori* solutions of the Maxwell equations (28), thus are not yet BT-conjugate solutions of the respective wave equations in (29). To find the restrictions these amplitudes must satisfy, we insert Eqs. (30) into the system (28). With the aid of the relation

$$\vec{\nabla} e^{\left(i - \frac{s}{k}\right)\vec{k}\cdot\vec{r}} = \left(i - \frac{s}{k}\right)\vec{k}\, e^{\left(i - \frac{s}{k}\right)\vec{k}\cdot\vec{r}} ,$$

it is not hard to show that (28a) and (28b) impose the conditions

$$\vec{k}\cdot\vec{E}_0 = 0 , \quad \vec{k}\cdot\vec{B}_0 = 0 \qquad (32)$$

Again, this means that the e/m wave is a transverse wave.

Substituting (30) into (28c) and (28d), we find two more conditions:

$$(k + is)\hat{\tau}\times\vec{E}_0 = \omega\vec{B}_0 \qquad (33)$$

$$(k + is)\hat{\tau}\times\vec{B}_0 = -(\varepsilon\mu\omega + i\mu\sigma)\vec{E}_0 \qquad (34)$$

However, (34) is not an independent equation since it can be reproduced by cross-multiplication of (33) by $\hat{\tau}$ and use of relations (31).

The BT-conjugate solutions of the wave equations (29) are now written:

$$\vec{E}(\vec{r},t) = \vec{E}_0\, e^{-s\hat{\tau}\cdot\vec{r}}\, e^{i(\vec{k}\cdot\vec{r}-\omega t)} ,$$
$$\vec{B}(\vec{r},t) = \frac{k + is}{\omega}(\hat{\tau}\times\vec{E}_0)\, e^{-s\hat{\tau}\cdot\vec{r}}\, e^{i(\vec{k}\cdot\vec{r}-\omega t)} \qquad (35)$$

To find the corresponding real solutions, we assume linear polarization of the e/m wave and set, as before,

$$\vec{E}_0 = \vec{E}_{0,R}\, e^{i\alpha} .$$

We also set

$$k + is = |k + is|e^{i\varphi} = \sqrt{k^2 + s^2}\, e^{i\varphi} ;$$
$$\tan\varphi = s/k .$$

Taking the real parts of Eqs. (35), we finally have:

$$\vec{E}(\vec{r},t) = \vec{E}_{0,R}\, e^{-s\hat{\tau}\cdot\vec{r}}\cos(\vec{k}\cdot\vec{r} - \omega t + \alpha) ,$$
$$\vec{B}(\vec{r},t) = \frac{\sqrt{k^2+s^2}}{\omega}(\hat{\tau}\times\vec{E}_{0,R})\, e^{-s\hat{\tau}\cdot\vec{r}}\cos(\vec{k}\cdot\vec{r} - \omega t + \alpha + \varphi) .$$

These results as well as those of the previous section are, of course, well known from classical electromagnetic theory. It is mathematically interesting, however, to revisit the problem of constructing solutions of Maxwell's equations from the point of view of Bäcklund transformations, treating the electric and the magnetic component of a plane e/m wave as BT-conjugate solutions.

## 6. Summary and concluding remarks

Bäcklund transformations (BTs) were originally devised as a tool for finding solutions of nonlinear partial differential equations (PDEs). They were later also proven useful as nonlocal recursion operators for constructing infinite sequences of symmetries and conservation laws of certain PDEs [2–7].

Generally speaking, a BT is a system of PDEs connecting two fields that are required to independently satisfy two respective PDEs in order for the system to be integrable for either field. If a solution of either PDE is known, then a solution of the other PDE is obtained by integrating the BT, without having to actually solve the latter PDE explicitly (which, presumably, would be a much harder task). In the case where the two PDEs are identical, an auto-BT produces new solutions of a PDE from old ones.

As described above, a BT is an auxiliary tool for finding solutions of a given (usually nonlinear) PDE, using known solutions of the same or another PDE. In this article, however, we approached the BT concept differently by actually inverting the problem. According to this scheme, it is the solutions of the BT itself that we are after, having parameter-dependent solutions of the PDEs that express the integrability conditions at hand. By a proper choice of the parameters, a pair of solutions of these PDEs may possibly be found that satisfies the given BT. These solutions are then said to be *conjugate* with respect to the BT.

A pedagogical paradigm for demonstrating this particular approach to the concept of a BT is offered by the Maxwell system of equations of Electromagnetism. We showed that this system can be thought of as a BT whose integrability conditions are the wave equations for the electric and the magnetic field. These wave equations have known, parameter-dependent solutions (monochromatic plane waves) with arbitrary amplitudes, frequencies, wave vectors, etc. By substituting these solutions into the BT, one may determine the required relations among the parameters in order that the plane waves also represent electromagnetic fields, i.e., are BT-conjugate solutions of the Maxwell system. The results arrived at by this method are, of course, well known in advanced Electrodynamics. The process of deriving them, however, is seen here in a new light by employing the concept of a BT.

We remark that the physical situation was examined from the point of view of a fixed inertial observer. Thus, since no spacetime transformations were involved, we used the classical form of the Maxwell equations (with $\vec{E}$ and $\vec{B}$ retaining their individual characters) rather than the manifestly covariant form of these equations.

An interesting conclusion is that the concept of a Bäcklund transformation, which has been proven extremely useful for finding solutions of nonlinear PDEs, can in certain cases also prove useful for integrating *linear systems* of PDEs. Such systems appear often in Physics and Electrical Engineering (see, e.g., [9]) and it would certainly be of interest to explore the possibility of using BT methods for their integration.

### Acknowledgment

I thank Aristidis N. Magoulas for many fruitful discussions.

### References

[1] C. J. Papachristou, *Symmetry and integrability of classical field equations*, http://arxiv.org/abs/0803.3688.

[2] C. J. Papachristou, *Potential symmetries for self-dual gauge fields*, Phys. Lett. A 145 (1990) 250.

[3] C. J. Papachristou, *Recursion operator and current algebras for the potential SL(N,C) self-dual Yang-Mills equation*, Phys. Lett. A 154 (1991) 29.

[4] C. J. Papachristou, *Lax pair, hidden symmetries, and infinite sequences of conserved currents for self-dual Yang-Mills fields*, J. Phys. A 24 (1991) L 1051.

[5] C. J. Papachristou, *Symmetry, conserved charges, and Lax representations of nonlinear field equations: A unified approach*, Electron. J. Theor. Phys. 7, No. 23 (2010) 1.

[6] C. J. Papachristou, B. K. Harrison, *Bäcklund-transformation-related recursion operators: Application to the self-dual Yang-Mills equation*, J. Nonlin. Math. Phys., Vol. 17, No. 1 (2010) 35.

[7] C. J. Papachristou, *Symmetry and integrability of a reduced, 3-dimensional self-dual gauge field model*, Electron. J. Theor. Phys. 9, No. 26 (2012) 119.

[8] D. J. Griffiths, *Introduction to Electrodynamics*, 3rd Edition (Prentice-Hall, 1999).

[9] E. C. Zachmanoglou, D. W. Thoe, *Introduction to Partial Differential Equations with Applications* (Dover, 1986).